\documentclass[a4paper,fleqn,usenatbib]{mnras}
\usepackage{times}
\usepackage[authoryear]{natbib}
\usepackage{url}
\usepackage{graphicx}
\usepackage{mathrsfs}  
\usepackage[fleqn]{amsmath}
\usepackage{aas_macros}
\usepackage{afterpage}
\usepackage{xparse}
\usepackage{float}
\usepackage{amsmath}
\usepackage{subfig}
\usepackage{caption}

\usepackage{enumitem}

\def \simless {\mathbin{\lower 3pt\hbox{$\rlap{\raise 4pt
              \hbox{$\char'074$}}\mathchar"7218$}}}
\def \simgreat {\mathbin{\lower 3pt\hbox{$\rlap{\raise 4pt
              \hbox{$\char'076$}}\mathchar"7218$}}}

\def\eg {{\it e.g.}}

\usepackage[usenames]{color}
\makeatletter

\newcommand{\Rmnum}[1]{\expandafter\@slowromancap\romannumeral #1@}
\makeatother
\usepackage{accents}
\usepackage{cite}
\usepackage{amsmath}

\title{Submillimeter flux as a probe of molecular ISM mass in high-z galaxies}
\author[L. Liang et al.]{Lichen Liang$^{1}$, Robert Feldmann$^{1}$, Claude-Andr\'{e} Faucher-Gigu\`{e}re$^{2}$, Du\v{s}an Kere\v{s}$^{3}$, 
\newauthor Philip F. Hopkins$^{4}$, Christopher C. Hayward$^{5}$, Eliot Quataert$^{6}$, Nick Z. Scoville$^{4}$  \\
\\
$^{1}$Institute for Computational Science, University of Zurich, Zurich CH-8057, Switzerland\\
$^{2}$Department of Physics and Astronomy and CIERA, Northwestern University, Evanston, IL 60208, USA\\
$^{3}$Department of Physics, Center for Astrophysics and Space Sciences, University of California at San Diego, La Jolla, CA 92093, USA\\
$^{4}$TAPIR, California Institute of Technology, Pasadena, CA, USA\\
$^{5}$Center for Computational Astrophysics, Flatiron Institute, 162 Fifth Avenue, New York, NY 10010, USA\\
$^{6}$Department of Astronomy, 501 Campbell Hall, University of California, Berkeley, CA, 94720, USA}

\begin{document}

\date{Accepted 2018 . Received 2018; in original form 2018}

\pagerange{\pageref{firstpage}--\pageref{lastpage}} \pubyear{2018}

\maketitle

\label{firstpage}

\begin{abstract}
Recent long wavelength observations on the thermal dust continuum suggest that the Rayleigh-Jeans (RJ) tail can be used as a time-efficient quantitative probe of the dust and ISM mass in high-$z$ galaxies. We use high-resolution cosmological simulations from the Feedback in Realistic Environment (FIRE) project to analyze the dust emission of $M_*\simgreat10^{10}\;M_{\odot}$ galaxies at $z=2-4$. Our simulations (\textsc{\small MassiveFIRE}) explicitly include various forms of stellar feedback, and they produce the stellar masses and star formation rates of high-$z$ galaxies in agreement with observations. Using radiative transfer modelling, we show that sub-millimeter (sub-mm) \textcolor{black}{luminosity} and molecular ISM mass are tightly correlated and that the overall normalization is in quantitative agreement with observations. Notably, sub-mm \textcolor{black}{luminosity traces} molecular ISM mass even during starburst episodes as dust \textcolor{black}{mass} and mass-weighted \textcolor{black}{temperature} evolve only moderately between $z=4$ and $z=2$, including during starbursts. 
Our finding supports the empirical approach of using \textcolor{black}{broadband} sub-mm flux as a proxy for molecular gas content in high-$z$ galaxies. We thus expect single-band sub-mm observations with ALMA to dramatically increase the sample size of high-$z$ galaxies with reliable ISM masses in the near future.
\vspace{-4pt}
\end{abstract}

\begin{keywords}
galaxies: evolution --- galaxies: high-redshift --- galaxies: ISM --- submillimetre: galaxies
\vspace{-30pt}
\end{keywords}

\section{Introduction}

\vspace{-10pt}
\noindent Molecular gas is a key component of the interstellar medium (ISM) affecting the physics of galaxy and star
formation. Measuring the molecular gas mass in galaxies in different environments and \textcolor{black}{at various evolutionary stages} provides important \textcolor{black}{constraints} to the nature of the relevant physical processes, such as gas heating and cooling, turbulence, and feedback \citep[see \eg][and references therein]{K16}. It is also essential for understanding a wide variety of observational properties and the scaling relationships of galaxies \citep[\eg][]{N07,T10,SL17}. 

Unfortunately, observing cold molecular hydrogen is extremely difficult, and CO rotational transition lines have widely been used to probe molecular ISM masses. While CO detections provide a handle on the physical state of the ISM, this method is time-intensive, and for high-$z$ galaxies, the bulk of the observations are limited to high-$J$ transitions. The conversion from a high-$J$ line intensity to a molecular ISM mass depends on the density and temperature of the CO-emitting gas, and could vary at the order of magnitude level between individual galaxies \citep[see \eg][]{S11,F12,CW13,N14,RS16,O16}. 
 
To overcome such disadvantages, there have been emerging efforts to use \textcolor{black}{alternative probes of the ISM mass in massive high-$z$ galaxies, such as the CI line emission} \citep[\eg][]{OB14,A18} \textcolor{black}{and the long-wavelength dust thermal emission} \citep[\eg][]{M12,S14,S16,H17,SL17,H18}. In particular, measuring ISM masses based on single-band \textcolor{black}{submm/mm} fluxes \textcolor{black}{from the dust emission} offers the prospect of \textcolor{black}{efficiently} generating large samples of high-$z$ galaxies with estimated ISM masses.

Recently, \citet[][hereafter S16]{S16} carried out an empirical calibration between the rest-frame specific luminosity at $850\rm \mu m$ ($L_{\rm 850 \mu m}$) and the CO-derived (via $J=1\rightarrow0$) molecular ISM mass ($M_{\rm mol}$) using a compiled sample of local star-forming spirals, ultra-luminous infrared galaxies (ULIRGs), and high-$z$ submillimetre galaxies (SMGs\footnote{SMGs are sub-mm sources with observed flux density at $850\rm \mu m$ ($S_{\rm 850\mu m}$) larger than a few mJy.}). They find that the $L_{\rm 850 \mu m}$-to-$M_{\rm mol}$ ratio of the galaxies in that sample is nearly constant with $L_{\rm 850 \mu m}$ suggesting that the flux density in the Rayleigh-Jeans (RJ) tail could be used as a proxy for $M_{\rm mol}$, \textcolor{black}{given the distance to the object}. 

Estimating the reliability of this approach is of paramount importance if sub-mm observations are to be used to infer physical properties of the baryonic cycle in high-redshift galaxies. One potential caveat with the high-$z$ data set of S16 is that it is limited to bright ($S_{\rm 850 \mu m}\simgreat1\;\rm mJy$) sources. It is thus unclear whether the $L_{\rm 850 \mu m}$-to-$M_{\rm mol}$ ratio applies to sub-mJy sources at high-$z$ that are observable with ALMA thanks to its high sensitivity and resolution \citep[\eg][]{O14,A16}. Furthermore, the molecular ISM masses in S16 are derived from CO line observations and are thus susceptible to various systematic uncertainties.

To assess the robustness of the sub-mm technique for estimating molecular ISM masses, we study the $M_{\rm mol}$-to-$L_{\rm 850 \mu m}$ conversion factor, $\alpha_{850}$, in massive ($M_*\simgreat10^{10}\;M_{\odot}$) high-$z$ ($z=2-4$) galaxies drawn from a suite of cosmological zoom-in simulations \citep[\textsc{\small MassiveFIRE},][]{F16a, F17} that are part of the Feedback In Realistic Environments (FIRE) project\footnote{\url{fire.northwestern.edu}} \citep{H14}.

A number of previous studies have greatly improved our understanding of the nature of the sub-mm sources, as well as how various galactic properties impact the sub-mm flux \citep[\eg][]{N10,H11,HN13,N15}. However, to our knowledge, there has not yet been any theoretical study that probes $\alpha_{850}$ with the help of 3D radiative transfer (RT) in a self-consistent cosmological setting. In addition, the present work improves on previous theoretical studies by increasing the numerical resolution and by modeling star formation and stellar feedback more realistically in a galactic context. 

This letter is organized as follows: In Section~\ref{Sec:2}, we present the simulation details and methodology of our analysis. We investigate the $\alpha_{850}$ conversion factor in Section~\ref{Sec:3}. We discuss caveats and conclude in Section~\ref{Sec:4}. 
Throughout this letter, we adopt cosmological parameters in agreement with the nine-year data from the Wilkinson Microwave Anisotropy Probe \citep{H13}, specifically $\Omega_{\rm m}=0.2821$, $\Omega_{\Lambda}=0.7179$, and $H_0=69.7\; \rm km\;s^{-1}\;Mpc^{-1}$.

\vspace{-24pt}
 
\section{Simulation and Analysis Methodology}
\label{Sec:2}

\vspace{-4pt}
%\begin{figure}
% \vspace{-6pt}
% \begin{center}
% \includegraphics[height=77mm,width=77mm]{figure1zb.pdf}
% \caption{{\it Top:} Rest-frame UVJ image of a selected \textsc{\small MassiveFIRE} galaxy at $z=2$, with (left) and without accounting for ISM dust (right). {\it Bottom:} Map of the intrinsic flux density at 850 $\rm \mu m$ (left) and surface mass density of molecular ISM (right). }
%    \label{fig.1}
%  \end{center}
%  \vspace{-11pt}
%\end{figure}

We extract our galaxy sample from the \textsc{\small MassiveFIRE} cosmological zoom-in suite \citep{F16a,F17}, {which implements all the same \textsc{FIRE} physics and numerical methods as other \textsc{FIRE-1} simulations \citep{H14}}. We include 18 massive ($10^{10}<M_*<10^{11.5}\;M_{\odot}$ at $z=2$) central galaxies (from the Series A, B and C in \citealt{F17}) and their most massive progenitors (MMPs), \textcolor{black}{identified using the Amiga Halo Finder (AHF) \citep{G04,K09}}. These galaxies reside in a variety of environmental overdensities and have a range of accretion histories. Masses and minimum adaptive gravitational softening lengths of individual gas (dark matter) particles are $3.3\times10^4\;M_{\odot}$ ($1.7\times10^5\;M_{\odot}$) and $\epsilon_{\rm b}=9$ (143) proper pc, respectively.  {FIRE simulations} account explicitly for stellar feedback and the multi-phase (molecular through hot) ISM \citep{H14}.  

We produce the UV-to-mm continuous SEDs for our \textsc{\small MassiveFIRE} sample from multiple viewing angles with the publicly available 3D Monte Carlo RT code \textsc{\small SKIRT}\footnote{\textsc{\small  SKIRT} accounts for the scattering, absorption and emission by dust, {and calculates the dust temperature iteratively until convergence}, see \url{www.SKIRT.ugent.be}.}\citep{B11,C15}, using the methodology of \citet{C16} (see also \citealt{T17})  except for the (here unnecessary) particle re-sampling. In brief, stars older than 10 Myrs are assigned tabulated SEDs according to their age and metallicity while younger stars are assigned \textsc{\small MappingsIII} source SEDs \citep{G08}. The PDR covering fraction is set to $f_{\rm pdr}=0.2$ \citep{G08,J10}. Dust associated with unresolved birthclouds of young star clusters is removed from their neighbouring gas particles to avoid double-counting. 

The \textsc{\small MappingsIII} SEDs account for the emission from the warm dust associated with the unresolved birthclouds of the young star clusters. {We hereafter refer to this dust component as `dense dust' and to the remaining interstellar component (which dominates the total dust mass) as `diffuse dust' throughout this letter. The temperature of the dense component is characterized by the `compactness' parameter ($C$) in the \textsc{MAPPINGSIII} model \citep[\textcolor{black}{see equation 13 of}][\textcolor{black}{for definition of $C$}]{G08}. For the high-$z$ results presented in this letter, we adopt ${\rm log}\;C=6.5$ following \citet{ND10}. Lowering $\rm log$ $C$ to 5.5 decreases the temperature of the diffuse dust component by $\simless5\%$.

 We adopt the Milky-Way dust size distribution from \citet{WD01} with {$\beta\approx2$} and a constant dust-to-metal mass ratio $M_{\rm dust}/M_{\rm Z}=0.4$ \citep{D98,DD07} in the cold {($T<10^6$ K)} ISM. For our sample, switching to an SMC or LMC dust model changes $L_{850\rm \mu m}$ typically by less than 0.06 dex. Hotter gas is assumed to be dust-free due to thermal sputtering \citep{HN15}. {We take into account dust self-absorption and re-emission in the analysis.} 

We use an octree dust grid and keep subdividing grid cells until the cell contains less than $f=3\times10^{-6}$ of the total dust mass {\it and} the V-band optical depth in each cell is smaller than unity. The highest grid level corresponds to a cell width of $\sim{}20$ pc, i.e., about twice the minimal SPH smoothing length. The RT calculations presented in this letter are converged {for the given resolution of our simulations.}

Unless noted otherwise, sub-mm \textcolor{black}{luminosity} and other properties of simulated galaxies are computed based on star and gas particles inside a $\rm (150\;kpc)^3$ box centred at each given galaxy. The chosen box size is comparable to the of the SCUBA beam size at 850$\rm \mu m$ at $z=2-4$. Molecular gas \textcolor{black}{mass} fractions of SPH particles are estimated using the local gas column density, {$\Sigma=\rho h=\rho\cdot \rho/|\nabla\rho|$}, and metallicity following the analytic prescription by \citet{KG11}.

\begin{figure}
 \vspace{-11pt}
 \begin{center}
 \includegraphics[height=59mm,width=77mm]{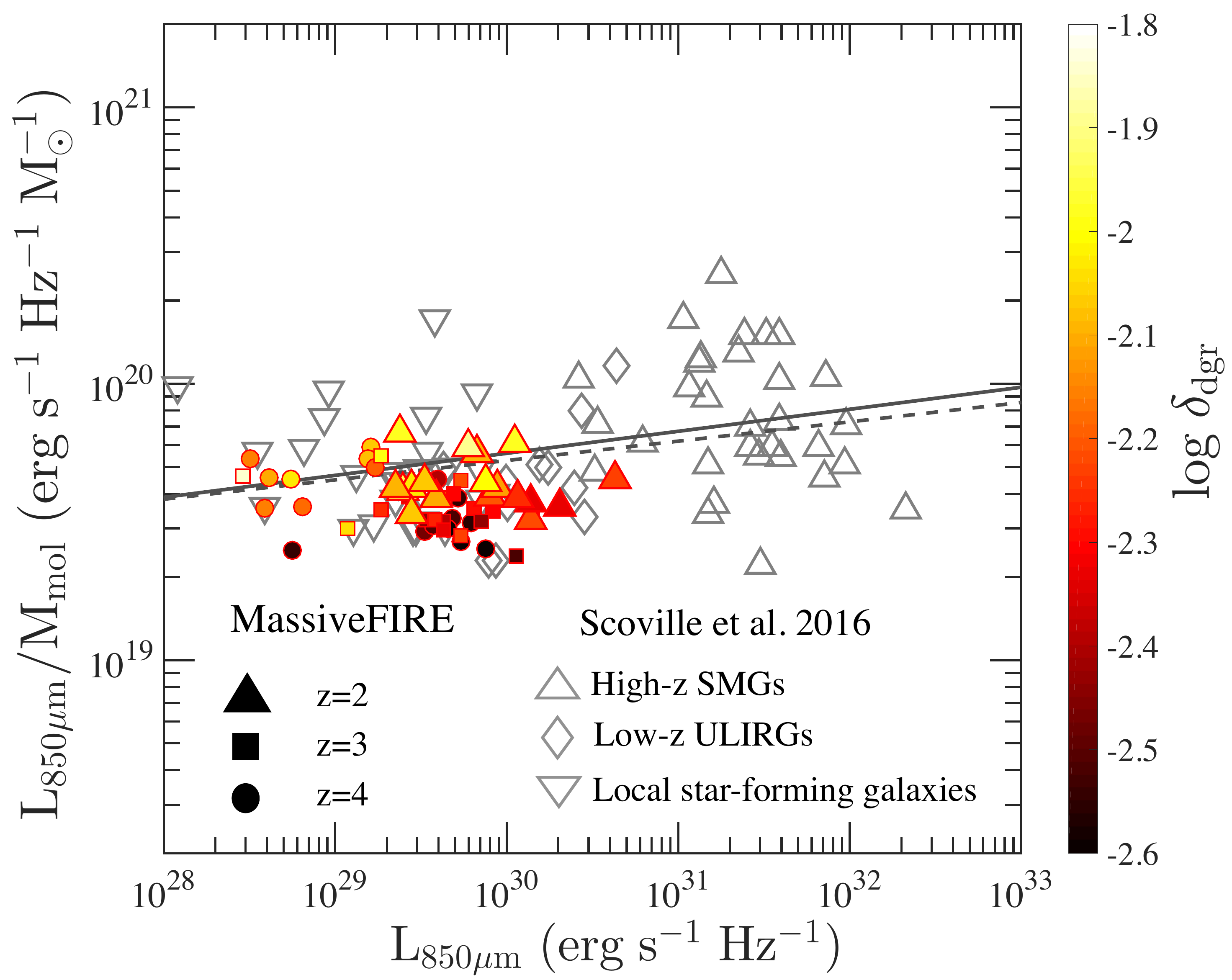}
 \vspace{-8pt}
 \caption{Ratio of $L_{\rm 850\mu m}$ to $M_{\rm mol}$, i.e., $1/\alpha_{850}$, as function of $L_{\rm 850\mu m}$. Filled symbols represent the result of \textsc{\small MassiveFIRE} galaxies at $z=2$ (triangles), $z=3$ (squares) and $z=4$ (circles), while empty symbols show the compilation of the recent observations from S16.  Empty triangles, diamonds, and inverse triangles represent SMGs, ULIRGs and local spirals, respectively. $L_{\rm 850\mu m}$ of SMGs are de-lensed by the previously-reported magnification factor $\mu$.  Grey lines show a linear fit to the S16 data. Dashed and solid line show results obtained using original and de-lensed $L_{\rm 850\mu m}$ of SMGs. Simulated galaxies are coloured according to their  dust-to-\textcolor{black}{(molecular)} gas \textcolor{black}{mass} ratio, $\delta_{\rm dgr}$, indicating a weak trend of $\alpha_{850}$ on $\delta_{\rm dgr}$. $\textsc{MassiveFIRE}$ simulations predict $\alpha_{850}$ values in good agreement with those of star forming and sub-mm galaxies at $z\sim0-2$. }
    \label{fig.1}
  \end{center}
  \vspace{-20pt}
\end{figure}

\vspace{-19pt}
\section{The results}
\label{Sec:3}
\vspace{-4pt}
\subsection{The sub-mm-to-CO scaling relation}
\label{Sec:3.1}
\vspace{-3pt}
%%%%%%%%%%%%%%%%%%%
%%More about Scoville's method %%
%%%%%%%%%%%%%%%%%%%

The S16 compilation includes 28 local star-forming spiral galaxies, 12 ULIRGs and 30 $z\sim1.5-3$ SMGs, with galaxy-integrated sub-mm dust emission and molecular ISM masses. S16 convert the observed SPIRE 500${\rm \mu m}$ (SCUBA 850$\rm \mu m$) flux of the local galaxies (high-$z$ SMGs) to the rest-frame specific luminosity $L_{\rm 850\mu m}$, using a single-temperature modified-blackbody spectra with mass-weighted dust temperature $T_{\rm dust}=25$ K and dust emissivity index $\beta=1.8$. The molecular ISM mass is derived from the {\it global} measurements of the integrated CO ($1-0$) fluxes with a single Galactic CO conversion factor $\alpha_{\rm CO}=M_{\rm mol}/L_{\rm CO}=6.5\;M_{\odot}/\rm K\;km\;s^{-1}pc^2$. Across the diverse samples of observed galaxies, the ISM-mass-to-sub-mm-luminosity ratio, $\alpha_{850}$, varies within a factor of 2 (see Fig.~\ref{fig.1}). In the same figure, we also show $\alpha_{850}$ for our sample of simulated MassiveFIRE galaxies at $z=2$, $z=3$, and $z=4$. \textcolor{black}{The $z=3$ and $z=4$ objects are the MMPs of the $z=2$ galaxies.}

The most luminous galaxy in our sample at $z=2$ has $S_{\rm 850\mu m}\approx\;1.5\;\rm mJy$ ($L_{\rm 850\mu m}\approx4.5\times10^{30}\rm \;erg\;s^{-1}\;Hz^{-1}$), which overlaps with (the fainter end of) observed $z\sim2$ SMGs. This galaxy is massive ($M_*\sim1.5\times10^{11}\;M_{\odot}$ within 0.1 $R_{\rm vir}$ excluding satellites) and undergoing multiple mergers at $z\sim{}2$. The stellar mass and \textcolor{black}{20-Myr-averaged} SFR within the SCUBA beam size centered on the galaxy are $\sim5\times10^{11}\;M_{\odot}$ and $\sim350\;M_{\odot}\;{\rm yr^{-1}}$. The conversion factor of this galaxy ($\alpha^{-1}_{\rm 850}\sim5\times10^{19}\;\rm erg\;s^{-1}\;Hz^{-1}\;M^{-1}_{\odot}$) is in good agreement with S16. The mass-weighted dust temperature of this system is roughly 27 K (as is shown in Fig.~\ref{fig.2}), similar to the dust temperature $T_{\rm dust}=25$ K adopted by S16.

The sample also contains 12 (9) candidates of faint SMGs at $z=2$ ($z=3$) with $S_{\rm 850\mu m}\simgreat 0.1\;\rm mJy$. One $z=4$ galaxy in our sample has $S_{\rm 1.3mm}\simgreat0.1\;\rm mJy$ \textcolor{black}{($z\sim4$ objects are often measured at longer wavelength to ensure that the observed sub-mm fluxes originate from the optically-thin RJ tail)}. The fainter $z=2$ \textsc{MassiveFIRE} galaxies are in good agreement with the linear scaling by S16, derived using local and high-$z$ samples. This suggests that the 850$\rm \mu m$ flux may be a reliable proxy for the molecular ISM mass of high-$z$ faint SMGs.

\begin{figure}
 \vspace{-11pt}
 \begin{center}
 \includegraphics[height=59mm,width=74mm]{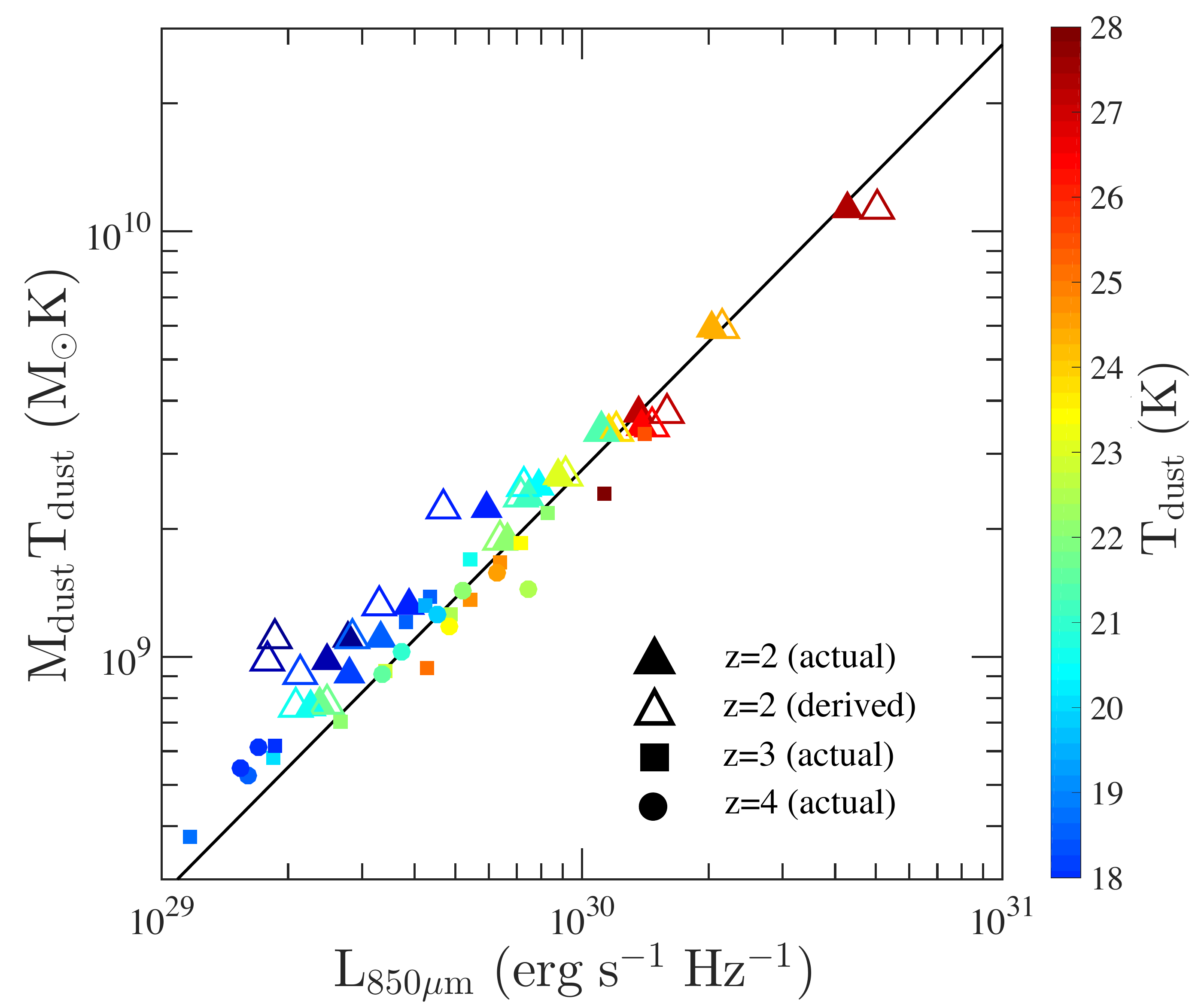}
  \vspace{-9pt}
 \caption{$M_{\rm dust} T_{\rm dust}$ against $L_{\rm 850 \mu m}$, where $M_{\rm dust}$ and $T_{\rm dust}$ represent the total mass and the mass-weighted temperature of the diffuse ISM dust. Triangles, squares and circles represent the \textsc{\small MassiveFIRE} galaxies at $z=2$, $z=3$ and $z=4$, respectively. For $z=2$, filled symbols show the actual $L_{\rm 850 \mu m}$, while empty symbols show $L_{\rm 850 \mu m}$ derived from $S_{\rm 850 \mu m}$ as in S16 with $T_{\rm dust}=25$ K. The diagonal line marks the linear scaling $M_{\rm dust}T_{\rm dust}=2.8\times10^{-21} L_{\rm 850\mu m}\rm \;M_{\odot}K\; s \;Hz\; erg^{-1}$. Data are coloured by $T_{\rm dust}$. The $850\rm \mu m$ luminosity is proportional to the dust mass and the mass-weighted temperature of the diffuse ISM dust.}
    \label{fig.2}
  \end{center}
   \vspace{-18pt}
\end{figure}

The data also reveals that $\alpha^{-1}_{\rm 850\mu m}$ depends on the  dust-to-\textcolor{black}{(molecular)} gas mass ratio, $\delta_{\rm dgr}$. Specifically, more dust-rich galaxies tend to have higher $\alpha^{-1}_{850\rm \mu m}$. As a consequence, $\alpha^{-1}_{850}$ increases by about 0.08 dex between $z=4$ to $z=2$ for the galaxies in our sample, mirroring their metal and dust enrichment.

We can understand the behavior of $\alpha_{850}$ qualitatively with the help of a single-temperature dust model. For such a model, the flux density in the RJ tail scales $\propto M_{\rm dust}T_{\rm dust}$ \citep{C14}. In Fig.~\ref{fig.2}, we show the scaling relationship between $M_{\rm dust}T_{\rm dust}$ and $L_{\rm 850\mu m}$ for the \textsc{\small MassiveFIRE} galaxies, where $M_{\rm dust}$ is the mass of the dust in the diffuse ISM and $T_{\rm dust}$ represents its mass-weighted temperature. It can be seen that $L_{\rm 850\mu m}$ is tightly correlated with $M_{\rm dust}T_{\rm dust}$ of the simulated galaxies, and therefore
\begin{equation}
\vspace{-10pt}
          \; \alpha^{-1}_{\rm 850 \mu m}=\frac{ L_{\rm 850\mu m}}{M_{\rm mol}} \propto \big ( \frac{M_{\rm dust}}{M_{\rm mol}}\big )T_{\rm dust} = \delta_{\rm dgr} \;T_{\rm dust}.
\label{eq.1}
\end{equation}

\noindent At $z=2$, our simulated galaxies have a approximately constant $\alpha_{850}$ due to their similar dust-to-gas \textcolor{black}{mass} ratios ($\delta_{\rm dgr}\approx1/125$) and dust temperatures ($T_{\rm dust}\sim23$ K). The gentle increase of $T_{\rm dust}$ with $L_{\rm 850 \mu m}$ seen in Fig.~\ref{fig.2} explains the minor increase of $\alpha^{-1}_{850}$ with increasing $L_{\rm 850 \mu m}$ at fixed $\delta_{\rm dgr}$ seen in Fig.~\ref{fig.1}. 

When using `derived' $L_{\rm 850\mu m}$ (see below) instead of actual $850\rm \mu m$ rest-frame \textcolor{black}{luminosities}, we find a weak correlation between $\alpha^{-1}_{850}$ and $L_{\rm 850\mu m}$ with slope $0.10\pm0.06$ for the full $z=2$ sample in good agreement with the empirical findings of S16. Derived $850\rm \mu m$ luminosities are computed based on $S_{\rm 850 \mu m}$ following the method by S16 (assuming $T_{\rm dust}=25$ K and $\beta=1.8$). We obtain $S_{\rm 850 \mu m}$ by convolving the redshifted SED with the transmission function of the SCUBA-2 850${\rm \mu m}$ filter \citep{HB13}. The derived $L_{\rm 850\mu m}$ is higher (lower) than the actual value for galaxies with $T_{\rm dust}$ higher (lower) than 25 K, as shown in Fig.~\ref{fig.2}.

Combining the data for all simulated galaxies with $S_{\rm 850\mu m}\simgreat0.1\;\rm mJy$ (for $z=2$ and $z=3$) and $S_{\rm 1.3mm}\simgreat0.1\;\rm mJy$ (for $z=4$), respectively, we obtain the following scaling relation\footnote{The quoted errors are standard errors of the linear regression and do not include systematic uncertainties.} for the conversion factor between molecular ISM mass and rest-frame $850\mu{}$m flux density
\vspace{-8pt}
\begin{equation}
\begin{split}
      \alpha^{-1}_{850} = 10^{(19.71\pm0.02)} \left[\frac{\delta_{\rm dgr}}{0.01}\right]^{(0.53\pm0.07)}&\left[\frac{L_{850\rm \mu m}}{10^{31}\;\rm erg/s/Hz} \right]^{(0.06 \pm 0.04)}\\
      &\rm erg\;s^{-1}\;Hz^{-1}\;M^{-1}_{\odot}.
\label{eq.2}
\end{split}
\end{equation} 
\vspace{-20pt}

\noindent The sub-unity of the scaling between $\alpha^{-1}_{850}$ and $\delta_{\rm dgr}$ results from a (non-trivial) anti-correlation between $T_{\rm dust}$ and $\delta_{\rm dgr}$ \citep[see also \eg][]{SH16}.

\begin{figure*}
 \vspace{-8pt}
 \begin{center}
 \includegraphics[height=75mm,width=138mm]{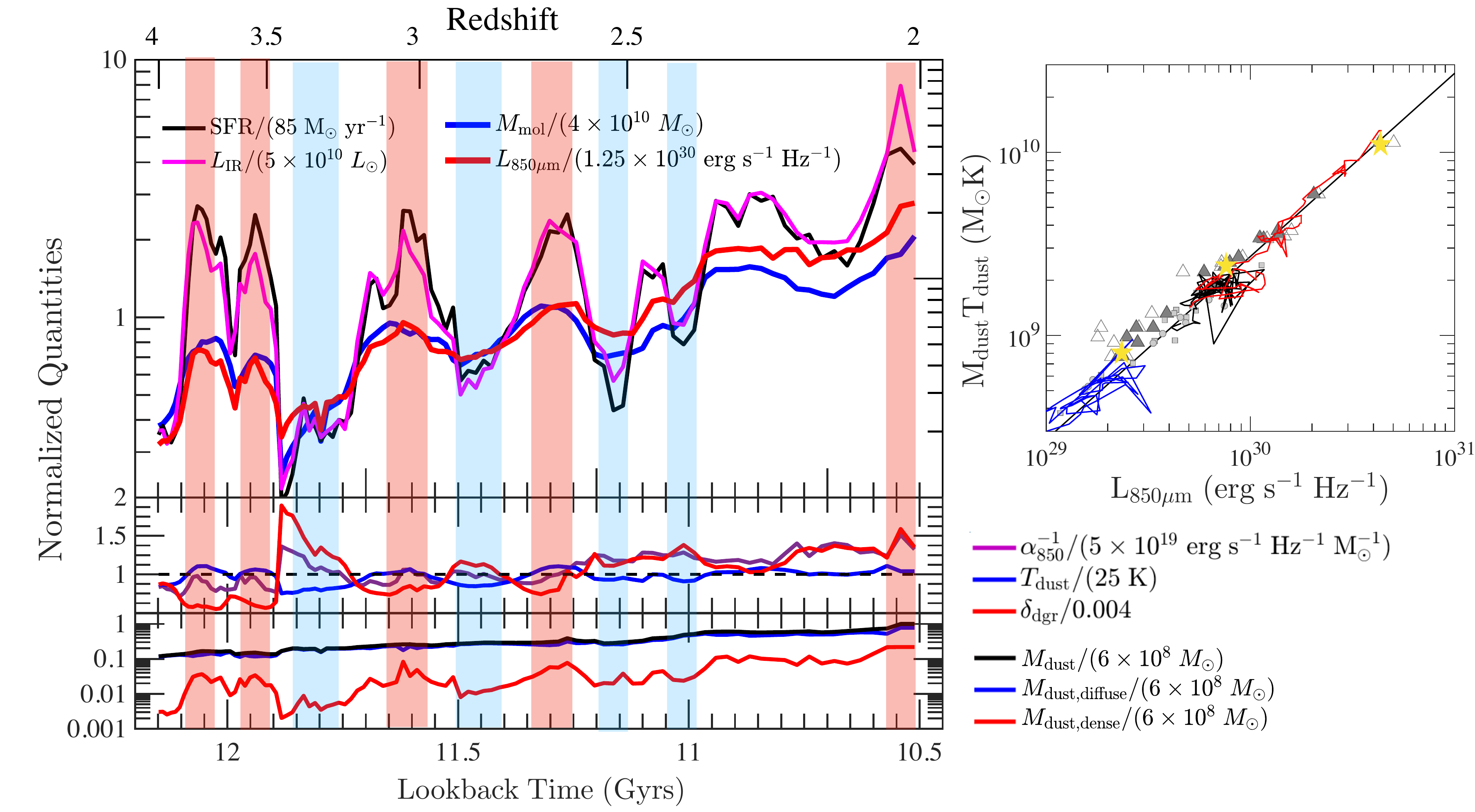}
 \vspace{-9pt}
 \caption{Temporal evolution of the various properties of the $L_{850\rm \mu m}$-brightest \textsc{\small MassiveFIRE} galaxy from $z=4$ to $z=2$. {\it Left:} In the top panel, the normalized $M_{\rm mol}$, SFR, $L_{\rm IR}$, and $L_{\rm 850 \mu m}$ are shown by the blue, black, magenta, and red lines. The middle panel shows the evolution of $L_{\rm 850 \mu m}/M_{\rm mol}$ (violet), $T_{\rm dust}$ (blue) and $\delta_{\rm dgr}$ (red). The mass of the dense (red line) and diffuse (blue line) dust are shown in the lower middle panel. Starburst and quiescent phases are marked by red and blue shaded blocks. {\it Right:} The subplot show the trajectories of three selected galaxies on the $M_{\rm dust} T_{\rm dust}$ vs. $L_{\rm 850\mu m}$ diagram. The red line corresponds to the $L_{850\rm \mu m}$-brightest galaxy, and black and blue lines show an intermediate and a low luminosity system. Yellow stars mark the position of these galaxies at $z=2$. $L_{850\rm \mu m}$ and $M_{\rm mol}$ co-evolve even during starbursts resulting in a roughly constant $\alpha_{850}$. $L_{\rm IR}$ traces well the SFR.}
    \label{fig.3}
  \end{center}
  \vspace{-19pt}
\end{figure*}

The scatter of the simulated $z=2$ data (0.09 dex) is smaller than that of the observed SMG sample (0.24 dex). The higher scatter in the observational data may arise from uncertainties in the CO ($1-0$) to $M_{\rm mol}$ conversion, which is known to depend on the ISM conditions of individual galaxies, including gas metallicity and internal UV radiation field \citep[\eg][]{G12,F12,B13,RS16}. Additionally, our simulations likely underestimate the scatter as they assume a fixed dust-to-metal ratio.

The PDR covering fraction, $f_{\rm pdr}$ appears to have little impact on $\alpha^{-1}_{850}$ if high-$z$ star forming regions are in a compact configuration (high $C$). Using $f_{\rm pdr}=1$ ($f_{\rm pdr}=0$), increases (decreases) $\alpha^{-1}_{850}$ by $\sim10\%$ ($1\%$).
Furthermore, $L_{\rm 850\mu m}/M_{\rm mol}$ does not vary significantly with the orientation of the galaxy because the long-wavelength dust continuum is optically thin.  

 \vspace{-16pt}
 
\subsection{The temporal evolution of sub-mm \textcolor{black}{luminosity}}

 \vspace{-5pt}
 
In the previous section, we studied the sub-mm \textcolor{black}{luminosity} and its relation to molecular ISM mass at fixed redshifts. We now investigate how the sub-mm \textcolor{black}{luminosity} of our simulated galaxies varies with time, especially during starburst (SB) episodes.

To this end, we track each $z=2$ galaxy back in time via their MMPs until $z=4$. For each galaxy we compute $L_{\rm 850\mu m}$ and calculate the total infrared luminosity, $L_{\rm IR}$ (over $\rm \lambda_{\rm rest}=8-1000\mu m$), using the output SED from \textsc{\small SKIRT}. We also calculate the molecular ISM mass,  20-Myr-averaged SFR, and total gas mass. As a representative example, we show in Fig.~\ref{fig.3} the temporal evolution of the various properties for the $L_{850\rm \mu m}$-brightest \textsc{MassiveFIRE} galaxy. 

This galaxy has several starbursting and quiescent phases on timescales of $\sim100$ Myrs. SBs are sometimes but not always associated with galaxy \textcolor{black}{mergers} in \textsc{MassiveFIRE} simulations \citep[][]{F17,SH17}. In most cases, SBs coincide with periods of increased gas density and molecular ISM mass. During a SB, molecular gas is quickly consumed by star formation, which subsequently drives gas away from the star forming region by stellar feedback \citep{M15} temporarily reducing SFRs \citep{F17}.

The figure demonstrates that $L_{\rm IR}$ is well correlated with the SFR. This is expected as $L_{\rm IR}$ traces the energy absorbed by the dense dust in vicinity of the young star clusters from their UV radiation, and is thus commonly used as an indicator of (obscured) star formation \citep[e.g.][]{K98b,C10}.

$L_{\rm 850\mu m}$, however, shows less correlation with SFR and instead mirrors the build-up of molecular ISM mass. How do we understand that $L_{850\mu{}m}$ tracks $M_{\rm mol}$ much better than $L_{\rm IR}$ even during SBs? In our simulations, $L_{\rm IR}$ arises largely from the hot dust close to the young star clusters, while $L_{\rm 850\mu m}$ traces the bulk of the (mostly diffuse) ISM dust. During the SBs, the mass fraction of the dense dust increases as the ISM becomes more compact, yet the total dust mass hardly changes and $T_{\rm dust}$ increases only slightly. As $L_{\rm 850\mu m}$ is well correlated with $M_{\rm dust}T_{\rm dust}$ (as shown in the right subplot of Fig.~\ref{fig.3}), $L_{\rm 850\mu m}$ tracks the (smoothly evolving) total dust mass and, consequently, $M_{\rm mol}$, while $L_{\rm IR}$ tracks the highly variable mass of dense and hot dust.
 
The reason for the weak evolution of dust temperature is twofold. First, during SB periods, gas and dust near the star-forming regions tend to be more compact resulting in increased absorption of UV radiation from embedded young star clusters. Re-emitted (IR) photons heat the remaining ISM dust inefficiently because of the low dust opacity at infrared wavelength. Secondly, the dust temperature scales only weakly (to the 1/6-1/5 power) with the local radiation density \citep{C14}. 

Hence, $\alpha^{-1}_{850}$ shows little short-term variance (changes are $\simless2\times$) even during SB episodes. Interestingly, modest deviations are sometimes seen during quiescent episodes when low molecular ISM masses drive temporary increases in $\alpha^{-1}_{850}$. Between $z=4$ and $z=2$, $\alpha^{-1}_{850}$ increases by $\sim{}60\%$ due to a growing $\delta_{\rm dgr}$. In contrast, over the same time frame, dust and gas masses as well as IR luminosities increase by one order of magnitude.

In summary, we find that $\rm 850\mu m$ flux density is a reliable proxy for molecular ISM mass even for starbursting systems.

\vspace{-24pt}
\section{Summary and Discussion}
\label{Sec:4}
\vspace{-4pt}
The long-wavelength dust thermal continuum has been used as an alternative to the traditional CO line method for constraining the molecular ISM mass of massive ($M_*\simgreat10^{10}\;M_{\odot}$) high-$z$ galaxies. We have coupled \textsc{\small MassiveFIRE} cosmological-zoom galaxies with dust radiative transfer to generate their UV-to-mm continuum SED. The sample covers two orders of magnitude in $L_{\rm 850 \mu m}$, with the brightest $z=2$ galaxy overlapping with the range of the observed SMGs. 

The ISM-mass-to-sub-mm-\textcolor{black}{luminosity}-ratio, $\alpha_{850}$, of our simulated sample of galaxies is in good agreement with the SMG sample from the S16 compilation (Fig.~\ref{fig.1}). Overall, $\alpha_{850}$ varies little with $L_{\rm 850 \mu m}$. $\alpha_{850}$ depends primarily on the dust temperature and  dust-to-(molecular) gas \textcolor{black}{mass} ratio and both quantities are similar among the galaxies in our sample.

When we look in more detail, we see that $\alpha_{850}$ slightly decreases with decreasing redshift as galaxies become successively more metal and dust-enriched. Specifically, we find $\alpha_{850}^{-1}=L_{\rm 850 \mu m}/M_{\rm mol}\propto \delta^{0.54}_{\rm dgr}$ (equation~\ref{eq.2}), which explains most of the scatter of $\alpha_{850}$ in our simulated sample. Indeed, the scatter in $\alpha_{850}$ at fixed $\delta_{\rm dgr}$ is small (0.02\;dex), showing that $\alpha_{850}$ largely tracks the dust enrichment in high-redshift galaxies. We therefore suggest the use of a \emph{dust-to-gas-ratio-dependent} conversion factor to increase the accuracy of the sub-mm-estimated ISM masses. {Observers can estimate $\delta_{\rm dgr}$ using the $\delta_{\rm dgr}-Z$ scaling \citep{M12}, and metallicities can be derived from \textcolor{black}{optical \citep[\eg][]{KE08} and FIR \citep[\eg][]{RP18}} line emission, or the $M_*-Z$ relationship \citep{T04,E06}.}

In our simulated galaxies, the mass-weighted dust temperatures ($T_{\rm dust}$) evolve relatively little between $z=4$ and $z=2$, rendering $\alpha_{850}$ insensitive to the burstiness of star formation.

Our findings thus suggest that single-band sub-mm techniques can reliably constrain the molecular ISM masses of both main-sequence and SB galaxies. Future single-band surveys with ALMA may dramatically increase the sample size of high-$z$ galaxies with well-constrained molecular ISM masses, and may thus provide important constraints on the physics of star formation and the gas cycling in galaxies at high redshift.

We thank the anonymous referee for comments that help improve the quality of this letter. We thank Onur \c{C}atmabacak, Tine Coleman, Nick Gnedin, Kevin Harrington, Claudia Lagos, Desika Narayanan for valuable discussions. We thank Peter Camps for providing SSP libraries for the \textsc{\small STARBURST99} stellar evolution model. Simulations were run with resources provided by the NASA High-End Computing (HEC) Program through the NASA Advanced Supercomputing (NAS) Division at Ames Research Center, proposal SMD-14-5492. Additional computing support was provided by HEC allocations SMD-14-5189, SMD-15-5950, and NSF XSEDE
allocations AST-120025, AST-150045. RF acknowledges financial support from the Swiss National Science Foundation (grant no 157591). CAFG was supported by NSF through grants AST-1412836, AST-1517491, AST-1715216, and CAREER award AST-1652522, and by NASA through grant NNX15AB22G. DK was supported by NSF grant AST-1715101 and the Cottrell Scholar Award from the Research Corporation for Science Advancement. EQ was supported in part by NSF grant AST-1715070 and a Simons Investigator Award from the Simons Foundation. The Flatiron Institute is supported by the Simons Foundation.
% [Put the following only there after the referee report.]
% We thank the referee for careful review and useful comments which help clarify the manuscript.
  \vspace{-18pt}

\bibliography{submm}{}
\bibliographystyle{mnras}
%\bibliographystyle{mnras}

%\bsp
% * <zjudracula@gmail.com> 2018-04-06T16:30:11.840Z:
%
% ^.
% * <zjudracula@gmail.com> 2018-04-06T16:30:09.796Z:
%
% ^.
\label{lastpage}
\end{document}